\documentclass[letterpaper, 10 pt, conference, onecolumn]{IEEEconf}  

\IEEEoverridecommandlockouts     
\usepackage{cite}
\usepackage{amsmath,amssymb,amsfonts}
\usepackage{algorithmic}
\usepackage{graphicx}
\usepackage{textcomp}

\usepackage{float} 
\usepackage{tcolorbox}
\usepackage{stfloats}
\usepackage{color}
\newtheorem{theorem}{\text{Theorem}}
\newtheorem{feature}{Feature}

\usepackage{tikz}
\usetikzlibrary{calc,patterns,decorations.pathmorphing,decorations.markings}
\usepackage{tcolorbox}%
\usepackage{preview}

\definecolor{darkgreen}{rgb}{0.1,0.8,0.5}
\definecolor{darkbrown}{rgb}{1.0, 0.25, 0.25}
\usetikzlibrary{positioning}
\usetikzlibrary{arrows}
\usepackage{subcaption}
\newsavebox{\tempbox}
\makeatletter
\let\NAT@parse\undefined
\makeatother
\usepackage[hidelinks]{hyperref}

\def\BibTeX{{\rm B\kern-.05em{\sc i\kern-.025em b}\kern-.08em
    T\kern-.1667em\lower.7ex\hbox{E}\kern-.125emX}}
\begin{document}
\title{Event-Triggered Distributed Stabilization of Interconnected Multiagent Systems with Abnormal Agent and Control Layers:\\Theoretical Analysis\vspace{-0.in}}
\author{Vahid Rezaei$^\star$\vspace{-0.5in}
\thanks{$^\star$V. Rezaei is with the Department of Computer Science and Engineering, University of Colorado, Denver, CO 80204, USA (Email: firstname.lastname@ucdenver.edu).} 
}
\maketitle

\thispagestyle{plain}
\pagestyle{plain} 
 
\begin{abstract}
A graph theoretic framework recently has been proposed to stabilize interconnected multiagent systems in a distributed fashion, while systematically capturing the architectural aspect of cyber-physical systems with separate agent or physical layer and control or cyber layer. Based on that development, in addition to the modeling uncertainties over the agent layer, we consider a scenario where the control layer is subject to the denial of service attacks. We propose a step-by-step procedure to design a control layer that, in the presence of the aforementioned abnormalities, guarantees a level of robustness and resiliency for the final two-layer interconnected multiagent system. The incorporation of an event-triggered strategy further ensures an effective use of the limited energy and communication resources over the control layer. We theoretically prove the resilient, robust, and Zeno-free convergence of all state trajectories to the origin and, via a simulation study, discuss the feasibility of the proposed ideas.
\end{abstract}

\section*{Introduction}
As a response to the advances in embedded communication, computation, and sensing technologies, multiagent systems (MASs) and cyber-physical systems (CPSs) are receiving significant attention among policymakers and researchers. These increasingly important systems are prone to various abnormalities over their physical (agent) and cyber (control) layers. By capturing the architectural aspect of CPSs, the following publication has tried to (at least partly) provide a foundation in order to systematically study the impact of cyber and physical abnormalities on the stability of interconnected MASs:\vspace{0.12in}

\begin{itemize}
	\item {\sl \emph{Rezaei V., ``Event-Triggered Distributed Stabilization of Interconnected Multiagent Systems with Abnormal Agent and Control Layers,"} IEEE Conference on Decision and Control, \emph{USA, Dec 2021.}}\vspace{0.12in}
\end{itemize}

In this brief, we provide a theoretical analysis for Theorem~1 of the above reference (i.e., main paper). Further details regarding the design steps, required definitions, parameters, variables, as well as further references are available in the main paper.

\section*{Overview and Theoretical Analysis}
We consider an interconnected MAS of $N$ fully heterogeneous agents:
\begin{equation}\label{eq:AgentModel}
	\begin{array}{rl}
		\dot{x}_i(t) & = A_i x_i(t) + B_{u_i} u_i(t) + B_{f_i}f_i(z_i(t),t)\\
		         z_i(t) & = C_{z_i} \sum_{j \in \mathcal{N}_i^a} a_{ij}^a x_j(t)
	\end{array}~~~\vspace{0.04in}
\end{equation}
where the parameters are defined in the main paper (entitled in Introduction).

Despite the modeling uncertainties over agent layer as well as DoS attacks and limited (energy) resources over control layer, the objective is to develop a control layer that guarantees the Zeno-free, exponential convergence of all state trajectories to the origin:
\begin{equation}\label{eq:Objective}
	\|x_i(t)\| \leq b_\star \exp^{-\sigma_\star t} \to 0 ~~~\text{as} ~~~ t \to \infty
\end{equation}
where $b_\star$ and $\sigma_\star$ are two positive scalars to be understood in the proof of Theorem~\ref{Theorem:Main}.

We propose the following distributed stabilization protocol in order to stabilize an interconnected MAS of agents~\eqref{eq:AgentModel}:
\begin{equation}\label{eq:DistributedStabilizationProtocol}
	u_i(t) = \sum_{j \in \mathcal{N}_i^c} a_{ij}^c (\hat{v}_i(t) - \hat{v}_j(t)) + s_i^c \hat{v}_i(t)
\end{equation}
where
\begin{equation}\label{eq:BufferAgenti}
	\hat{v}_i(t) = v_i(t_k^i)~~~~~\forall t \in [t_k^i,t_{k+1}^i)
\end{equation}
and
\begin{equation}\label{eq:BufferAgentj}
	\hat{v}_j(t) = v_j(t_k^j) ~~~~~ \forall t \in [t_k^j,t_{k+1}^j), ~~~~j \in \mathcal{N}_i^c.
\end{equation}
are obtained using the following virtual stabilization signal associated to each agent:
\begin{equation*}
	v_i(t) = K_i x_i(t).
\end{equation*}

The $i^{th}$ agent's information broadcast time sequence $\{t_k^i\}$ is automatically generated according to the following hybrid (mixed event-triggered and periodic information broadcast) strategy:
\begin{equation}\label{eq:ETS}
	t_{k+1}^i = \begin{cases}
		s_{k+1}^i, ~~~~~~~~~~~~ t \notin H_n\\
		t_k^i+t_{dos}, ~~~~~~~~ t \in H_n
	\end{cases}~~~~~ n \geq 0 \vspace{0.02in}
\end{equation}
in which, in the absence of DoS, each agent's broadcast of information is determined according to the following nonperiodic (agent-wise) and nonsynchronous (MAS-wise) ETS:
\begin{IEEEeqnarray*}{rll}
		& s_{k+1}^i = \inf \{t > s_{k}^i,~ t_0^i = 0~|~ \phi_i(t,x_i,\hat{x}_i) \leq 0\} \\
		& \phi_i(t,x_i,\hat{x}_i) = \kappa_{1i} \exp^{-\sigma t} + \kappa_{2i} \|x_i\|^2 - \|e_{vi}\|^2. \nonumber
\end{IEEEeqnarray*}

Consequently, we find the following model for the agent layer:
\begin{equation}\label{eq:AgentLayer}
\begin{array}{rl}
	\dot{x} & = \bar{A} x + \bar{B}_u u + \bar{B}_f f(z)\\
			z & = \bar{C}_z (\mathcal{A}_a \otimes I_{n_x}) x
\end{array}
\end{equation}
and for the control layer:
\begin{equation}\label{eq:ControlLayer}
	u = (\mathcal{H}_c \otimes I_{n_u}) \hat{v} = (\mathcal{H}_c \otimes I_{n_u}) \bar{K} \hat{x} = (\mathcal{H}_c \otimes I_{n_u}) \bar{K} (x + e).
\end{equation}
Also, the two-layer interconnected MAS would be as follows:
\begin{IEEEeqnarray}{rll}\label{eq:TwoLayerMAS}
\begin{array}{rl}
	\dot{x} = \bar{A} x + \mu_{c1} \bar{B}_u v + \mu_{c1} \bar{B}_u \bar{E}_c v + \bar{B}_u (\mathcal{H}_c \otimes I_{n_u}) e_v + \bar{B}_f f(z)
\end{array}~~~
\end{IEEEeqnarray}
where $\bar{E}_c = \big((\frac{\mathcal{H}_c}{\mu_{c1}} - I_N) \otimes I_{n_u}\big) \succcurlyeq \textbf{0}$. 

DoS attacks happen at time sequences $\{h_n\}$ with $\tau_n$ as its $n^{th}$ attack duration:
\begin{equation}
	H_n := \{h_{n}\} \cup [h_n, h_n + \tau_n).
\end{equation}

We let $\Xi(\tau,t)$ be an accumulative DoS time interval on $[\tau,t]$, and $\Theta(\tau,t)$ the total DoS-free interval:
\begin{equation}
	\begin{array}{rl}
		& \Xi(\tau,t) = \bigcup\limits_{n \geq 0} H_n \bigcap [\tau,t]\\
		& \Theta(\tau,t) = [\tau,t] \backslash \Xi(\tau,t)
	\end{array}~~~~~~~~~~~~~~~~\forall t \geq \tau \geq 0.
\end{equation}

Also, let $n(\tau,t)$ be the number of DoS off-to-on transitions during $[\tau,t]$. Now a class of DoS attacks can be characterized by the following two features.\vspace{0.05in}
\begin{feature}\label{assumption:DoSFrequency}(DoS frequency)
	The following inequality holds:
	\begin{equation}
		n(\tau,t) \leq \pi_f + \frac{t-\tau}{\tau_f}~~~~~~~\forall t \geq \tau \geq 0 \nonumber
	\end{equation}
	for some $\pi_f \geq 0$ and $\tau_f >0$.\vspace{0.05in}\hfill $\blacktriangleleft$
\end{feature}
\begin{feature}\label{assumption:DoSDuration}(DoS duration)
	The following inequality holds:
	\begin{equation}
		|\Xi(\tau,t)| \leq \pi_d + \frac{t-\tau}{\tau_d}~~~~~~~\forall t \geq \tau \geq 0 \nonumber
	\end{equation}
	for some $\pi_d \geq 0$ and $\tau_d > 1$.\vspace{0.05in}\hfill $\blacktriangleleft$
\end{feature}

If we follow the steps of Design Procedure 1 (see the main paper), we find a valid control layer if a validation matrix $\bar{Q}_v$ satisfies the following (sufficient) condition:
\begin{equation}\label{eq:UM:DP:SufficientCondition} 
	\bar{Q}_v \succ \textbf{0}.
\end{equation}	
Based on the solutions to the following algebraic Riccati equations:
\begin{equation}\label{eq:ARE}
	A_i^T P_i + P_i A_i + W_{xi} - \mu_{c1}^2 P_i B_{u_i} W_{vi}^{-1} B_{u_i}^T P_i = \textbf{0}
\end{equation}
we know that the distribution stabilization gains can be characterized as follows:
\begin{equation}\label{eq:ControlGain}
	K_i = - \mu_{c1} W_{vi}^{-1} B_{u_i}^T P_i.
\end{equation}
After a few manipulations, these latter equalities end in the following design properties:\vspace{0.05in}
	\begin{equation*}
		\begin{array}{rl}
	     	x^T \bar{W}_x x + v^T \bar{W}_v v + \bar{V}_x^T (\bar{A} x + \mu_{c1} \bar{B}_u v) = & 0\\
	     	2 v^T \bar{W}_v + \mu_{c1} \bar{V}_x^T \bar{B}_u = & \textbf{0}.
		\end{array}
	\end{equation*}

We further define:
\begin{equation*}
	\begin{array}{rl}
	 	\rho_{dos} & = \frac{4 \max_i\{\lambda_{max}(K_i^T K_i)\}}{a_e \lambda_{min}(\bar{P})} ~~~~ \text{and} ~~~~ \frac{1}{\tau_\star} = \frac{1}{\tau_d} + \frac{t_{dos}}{\tau_f}
	\end{array}
\end{equation*} 
for all $i \in \{1,2,...,N\}$, and $\tau_f$ and $\tau_d$ in Features~1 and~2. Now we are ready to provide a proof for Theorem~1 in the main paper (mentioned in Introduction).\vspace{0.15in}

\begin{theorem}\label{Theorem:Main}
	Based on a two-layer interconnected MAS~\eqref{eq:TwoLayerMAS}:
	\begin{enumerate}
		\item In the absence of DoS, all state trajectories exponentially converge to the origin.
		\item In the presence of DoS, all state trajectories converge to the origin if the following condition is satisfied:
			\begin{equation}\label{eq:UM:DoS:Thm:Condition}
				 \frac{\rho_v - \sigma}{\rho_v + \rho_{dos}} < \frac{1}{\tau_\star}< \frac{\rho_v}{\rho_v + \rho_{dos}}.
			\end{equation}
		\item The Zeno phenomenon is ruled out.\vspace{0.05in}
    \end{enumerate}	
\end{theorem}
\begin{proof}
We prove this theorem in three parts:\vspace{0.05in}

\noindent \underline{\textit{(Part~1)}} We propose the following positive definite (candidate Lyapunov) function to prove the robust exponential convergence of all trajectories to the origin:
\begin{equation*}
	\bar{V}(x) = x^T \bar{P} x \succ 0.
\end{equation*}
Along the (uncertain) trajectories of \eqref{eq:TwoLayerMAS}, we find:
\begin{equation*}
	\begin{array}{rl}
		\dot{\bar{V}}(x) 
							  & = \bar{V}_x^T \big( \bar{A} x + \mu_{c1} \bar{B}_u v + \mu_{c1} \bar{B}_u \bar{E}_c v + \bar{B}_u (\mathcal{H}_c \otimes I_{n_u}) e_v + \bar{B}_f f\big) \\
							  & = \bar{V}_x^T \big( \bar{A} x + \mu_{c1} \bar{B}_u v \big) + \mu_{c1} \bar{V}_x^T \bar{B}_u \bar{E}_c v + \mu_{c1} \bar{V}_x^T \bar{B}_u (\frac{\mathcal{H}_c}{\mu_{c1}} \otimes I_{n_u}) e_v + \bar{V}_x^T  \bar{B}_f f \\
							  & = -x^T \bar{W}_x x - v^T \bar{W}_v v - 2v^T \bar{W}_v \bar{E}_c v - 2v^T \bar{W}_v (\frac{\mathcal{H}_c}{\mu_{c1}} \otimes I_{n_u}) e_v + 2 x^T \bar{P} \bar{B}_f f \\
							  & \leq -x^T \bar{W}_x x - v^T \bar{W}_v v - 2v^T \bar{W}_v \bar{E}_c v + a_e v^T \bar{W}_v (\frac{\mathcal{H}_c^2}{\mu_{c1}^2} \otimes I_{n_u}) \bar{W}_v v + a_f f^T f + \frac{1}{a_e} e_v^T e_v + \frac{1}{a_f} x^T \bar{P} \bar{B}_f \bar{B}_f^T \bar{P} x\\
							  & \leq -x^T (\bar{Q} + \frac{\kappa_2}{a_e}I_{Nn_x})x - v^T \bar{W}_v v - 2v^T \bar{W}_v \bar{E}_c v + a_e v^T \bar{W}_v (\frac{\mathcal{H}_c^2}{\mu_{c1}^2} \otimes I_{n_u}) \bar{W}_v v + \frac{1}{a_e} e_v^T e_v + \frac{1}{a_f} x^T \bar{P} \bar{B}_f \bar{B}_f^T \bar{P} x
	\end{array}
\end{equation*}
where we have used Design Properties~1 (see the main paper) and Young's inequality to obtain the first inequality, and the definitions of matrices in Design Procedure~1 (see the main paper) to obtain the second inequality. In particular, we reach to:
\begin{equation}\label{eq:UM:Theorem:LyapunovInequality4DoS}
	\begin{array}{rl}
		\dot{\bar{V}}(x) & \leq - x^T (\bar{Q}_v + \frac{\kappa_2}{a_e} I_{Nn_x} ) x + \frac{1}{a_e} \sum_{i = 1}^N \|e_{vi}\|^2
	\end{array}
\end{equation}
to be rewritten as follows using ETS~\eqref{eq:ETS} in its DoS-free case:
\begin{equation}\label{eq:UM:UpperBoundOnDotV}
	\begin{array}{rl}
		\dot{\bar{V}}(x) & \leq - x^T (\bar{Q}_v + \frac{\kappa_2}{a_e} I_{Nn_x}) x + \frac{1}{a_e} \sum_{i = 1}^{N} (\kappa_{1i} \exp^{-\sigma t} + \kappa_{2i} \|x_i\|^2) \\
						  	  & \leq -\rho_v \bar{V}(x) + \frac{1}{a_e} \sum_{i = 1}^{N} \kappa_{1i} \exp^{-\sigma t}
	\end{array}
\end{equation}
where $\rho_v$ is defined prior to Design Procedure~1 in the main paper. Now, based on the comparison lemma \cite{Khalil-Prentice-2003}, we find:
\begin{IEEEeqnarray}{rl}\label{eq:UM:UpperBoundOnV}
\begin{array}{rl}
	\bar{V}(t) 
	& \leq \exp^{-\rho_v t} \bar{V}(0) + \frac{1}{a_e} \sum_{i = 1}^N \frac{\kappa_{1i}}{\rho_v - \sigma} (\exp^{-\sigma t} - \exp^{-\rho_v t})~~~~
\end{array}
\end{IEEEeqnarray}
where we have introduced $\bar{V}(t) = \bar{V}(x(t))$. As a result, based on $0 < \sigma < \rho_v$ which holds by definition (see the setup description for ETS~\eqref{eq:ETS}), we know $ \|x_i(t)\|^2 \leq \|x(t)\|^2 \leq \frac{\lambda_{max}(\bar{P})}{\lambda_{min}(\bar{P})} \|x(0)\|^2 \exp^{-\rho_v t} + \frac{1}{a_e \lambda_{min}(\bar{P})} \sum_{i = 1}^N \frac{\kappa_{1i}}{\rho_v - \sigma} (\exp^{-\sigma t} - \exp^{-\rho_v t})$ which would end in:
\begin{equation}\label{eq:UM:UpperBoundonNormX_iSquare}
\begin{array}{rl}
	\|x_i(t)\|^2 
				   & \leq b_1 \exp^{-\sigma t}\\
%
\end{array}
\end{equation}
in which we have defined $b_1 := \frac{\lambda_{max}(\bar{P})}{\lambda_{min}(\bar{P})} \|x(0)\|^2 + \frac{1}{a_e \lambda_{min}(\bar{P})} \sum_{i = 1}^N \frac{\kappa_{1i}}{\rho_v - \sigma}$, and ignored the negative term associated to $- \exp^{-\rho_v t}$. This inequality is sufficient to conclude $x_i(t) \to \textbf{0}$ as $t \to \infty$ with a guaranteed exponential rate $\sigma > 0$.\vspace{0.1in}


\noindent \underline{\textit{(Part~2)}} Starting from \eqref{eq:UM:Theorem:LyapunovInequality4DoS}, we further find: 
\begin{IEEEeqnarray}{rll}\label{eqUMDoS:TheoremSETS_LyapunoveInequality}
\begin{array}{rl}
	\dot{\bar{V}}
					  & \leq \sum_{i = 1}^N \big(- \sqrt{\frac{\kappa_2 + a_e \lambda_{min}(\bar{Q}_v)}{a_e}} \|x_i(t)\|^2 + \frac{1}{a_e} \|K_i e_i(t)\|^2 \big) ~~~~~
\end{array}	
\end{IEEEeqnarray}
which remains valid either in the absence or in the presence of DoS. We assume a worst-case scenario where the entire communication network goes down in the presence of a DoS over the control layer, and agents must rely on only the last available information of the neighbors (in the associated buffers). Let $t_{dos,i}^s$ be the most recent successful triggering time of agent $i$ which has happened prior to DoS. During this attack, we have:
\begin{equation}\label{eqDoS:Buffers}
	\hat{x}_i(t) = x_i(t_{dos,i}^s)~~~~~~~~~~\forall t \geq t_{dos,i}^s.
\end{equation}
Now, based on the above foundation, we divide this proof into three subparts:\vspace{0.04in}

\noindent \underline{\textit{(Subpart~2.1 - DoS-free interval $[h_{n}+\tau_{n} + t_{dos}, h_{n+1})$)}} Prior to time $t_{dos,i}^s$, the two-layer interconnected MAS operates in its normal mode under ETS~\eqref{eq:ETS}. We integrate both sides of~\eqref{eq:UM:UpperBoundOnDotV} over $[h_{n}+\tau_{n} + t_{dos}, t)$, and use comparison lemma to find:
\begin{equation*}
	\begin{array}{rl}
		\bar{V}(t) 
					  & \leq \exp^{-\rho_v \big(t - (h_{n}+\tau_n + t_{dos})\big)} \bar{V}(h_{n}+\tau_n + t_{dos}) + b_2 \exp^{-\sigma t}
	\end{array}
\end{equation*}
where $b_2 = \frac{1}{a_e} \sum_{i = 1}^N \frac{\kappa_{1i}}{\rho_v - \sigma}$, and we have ignored the negative term associated to $- \exp^{(\rho_v - \sigma) (h_n + \tau_n)} \exp^{-\rho_v t}$ in order to obtain the right hand-side of this inequality.\vspace{0.04in}

\noindent \underline{\textit{(Subpart~2.2 - DoS interval $[h_n,h_n +\tau_n + t_{dos})$)}} In this case, the triggering error $e_i$ of agent $i$ is as follows:
\begin{equation}
	e_i(t) = \hat{x}_i (t_{dos,i}^s) - x_i(t)~~~~~~~\forall t \geq t_{dos,i}^s.
\end{equation}
Accordingly, we know $\|e_i(t)\|^2 \leq 2 \|\hat{x}_i(t_{dos,i}^s) \|^2 + 2 \| x_i(t)\|^2$ and find:
\begin{equation*}
   \sum_{i = 1}^N \|K_i e_i(t)\|^2 \leq 2 \sum_{i = 1}^N \|K_i \hat{x}_i(t_{dos,i}^s) \|^2 + 2 \sum_{i = 1}^N \|K_i x_i(t)\|^2.
\end{equation*}
Two cases may arise:
\begin{enumerate}
	\item $\sum_{i = 1}^N \| K_i \hat{x}_i(t_{dos,i}^s) \|^2 \leq \sum_{i = 1}^N \| K_i x_i(t)\|^2$ which results in:
	\begin{equation*}
		\sum_{i = 1}^N\|K_i e_i(t)\|^2 \leq 4 \sum_{i = 1}^N \|K_i x_i(t)\|^2.
	\end{equation*}
	We proceed with inequality~\eqref{eqUMDoS:TheoremSETS_LyapunoveInequality}, and find:
	\begin{equation}\label{eq:UM:DoS:TheoremSETS_LyapunovInequality1}
	\begin{array}{rl}
		\dot{\bar{V}}(t) \leq & \sum_{i = 1}^N \big(- \sqrt{\frac{\kappa_2 + a_e \lambda_{min}(\bar{Q}_v)}{a_e}} \| x_i(t)\|^2 + \frac{4}{a_e}  \| K_i x_i(t)\|^2 \big)\\
				   		 \leq & \sum_{i = 1}^N \big( \frac{4 \lambda_{max}(K_i^T K_i)}{a_e} - \sqrt{\frac{\kappa_2 + a_e \lambda_{min}(\bar{Q}_v)}{a_e}} \big) \|x_i(t)\|^2\\
				   		 \leq & \rho_{dos} \sum_{i = 1}^N x_i^T(t) P_i x_i(t) =: \rho_{dos} \bar{V}(x(t))
	\end{array}
	\end{equation}
	where $\rho_{dos}$ is defined prior to the main statement of this theorem.
	
	\item $\sum_{i = 1}^N \| K_i x_i(t)\|^2 \leq \sum_{i = 1}^N \| K_i \hat{x}_i(t_{dos,i}^s) \|^2$ which results in:
	\begin{equation*}
		\sum_{i = 1}^N\| K_i e_i(t)\|^2 \leq 4 \sum_{i = 1}^N \| K_i \hat{x}_i(t_{dos,i}^s) \|^2.
	\end{equation*}
	We proceed with inequality~\eqref{eqUMDoS:TheoremSETS_LyapunoveInequality}, and find:
	\begin{equation}\label{eq:UM:DoS:TheoremSETS_LyapunovInequality2}
	\begin{array}{rl}
		\dot{\bar{V}}(t) \leq & \sum_{i = 1}^N \big(- \sqrt{\frac{\kappa_2 + a_e \lambda_{min}(\bar{Q}_v)}{a_e}} \| x_i(t)\|^2  + \frac{4}{a_e} \sum_{i = 1}^N \| K_i \hat{x}_i(t_{dos,i}^s) \|^2 \big) \\
							  \leq & \rho_{dos} \sum_{i = 1}^N x_i^T(h_n) P_i x_i(h_n)=: \rho_{dos} \bar{V}(h_n).
		\end{array}
	\end{equation}
\end{enumerate}

\noindent Based on \eqref{eq:UM:DoS:TheoremSETS_LyapunovInequality1} and \eqref{eq:UM:DoS:TheoremSETS_LyapunovInequality2}, we reach to the following inequality:
\begin{equation}
	\dot{\bar{V}}(t) \leq \rho_{dos} \max\{ \bar{V}(t), \bar{V}(h_n) \} ~~~~~~~~ \forall t \in [h_n, h_n + \tau_n + t_{dos}).
\end{equation}
We focus on $\bar{V}(t) \geq \bar{V}(h_n)$ for $t \in [h_n, h_n + \tau_n +t_{dos})$, because $\bar{V}(t) \leq \bar{V}(h_n)$ would be trivial as it means DoS does not have any sever (divergent) impact on the underlying interconnected MAS. Thus, we find
\begin{equation*}
	\dot{\bar{V}}(t) \leq \rho_{dos} \bar{V}(t) ~~~~~~~~~~ \forall t \in [h_n, h_n + \tau_n + t_{dos})
\end{equation*}
and, consequently,
\begin{equation}\label{eq:StaticDoSRho_V1}
	\bar{V}(t) \leq \exp^{\rho_{dos} (t - h_n)} \bar{V}(h_n)~~~~~~~~~~ \forall t \in [h_n, h_n + \tau_n + t_{dos}).
\end{equation} \vspace{0.025in}
\noindent \underline{\textit{(Subpart~2.3 - The entire time $[0,t)$)}} Now we integrate the results of Subparts 2.1 and 2.2. In particular,
\begin{equation}\label{eqStaticDos:FinalBound1}
	\begin{array}{rl}
		\bar{V}(t) & \leq \exp^{-\rho_v \big(t - (h_{n}+\tau_n + t_{dos})\big)} \bar{V}(h_n+\tau_n + t_{dos}) + b_2 \exp^{-\sigma t}\\
					  & \leq \exp^{-\rho_v \big(t - (h_{n}+\tau_n + t_{dos})\big)} \exp^{\rho_{dos} (\tau_n + t_{dos})} \bar{V}(h_n) + b_2 \exp^{-\sigma t}\\
	& \leq \exp^{-\rho_v\big(t - \sum_{q = n}^{n-1} (\tau_q + t_{dos})\big)} \exp^{\rho_v h_{n-1}}  \exp^{\rho_{dos} (\tau_n + t_{dos})}  \\
	& ~~~~~ \times \bar{V}(h_{n-1}+\tau_{n-1} + t_{dos}) + b_2 \exp^{-\rho_v \big(t - (h_{n}+\tau_n + t_{dos})\big)} \\
	& ~~~~~ \times \exp^{\rho_{dos} (\tau_n + t_{dos})} \exp^{-\sigma h_n} + b_2 \exp^{-\sigma t}
	\end{array} 			
\end{equation}

\noindent which, eventually, would end in the followings:
\begin{equation}\label{eqM:DoS:Thm:FinalV}
	\begin{array}{rl}
		\bar{V}(t) \leq & \exp^{-\rho_v \bar{\Theta}(0,t) + \rho_{dos} \bar{\Xi}(0,t) } \bar{V}(0) + b_2 \exp^{-\sigma t}\\
							& + b_2'(h_1) \exp^{-\rho_v \bar{\Theta}(h_1,t) + \rho_{dos} \bar{\Xi}(h_1,t) }\\
							& \vdots\\
							& + b_2'(h_{n-1}) \exp^{-\rho_v \bar{\Theta}(h_{n-1},t) + \rho_{dos} \bar{\Xi}(h_{n-1},t) }\\
							& + b_2'(h_n) \exp^{-\rho_v \bar{\Theta}(h_{n},t) + \rho_{dos} \bar{\Xi}(h_{n},t) }
	\end{array} 			
\end{equation}
where $b_2'(h_m) = b_2 \exp^{(\rho_v - \sigma) h_m}$ for $m \in \{1,2,..,n\}$, $\bar{\Xi}(h_m,t)$ denotes the total time interval during which the communication is blocked (including an additional time period $t_{dos}$ after each DoS interval $H_q$ and before the next ETS-based information broadcast), and $\bar{\Theta}(h_m,t)$ that of free communication over the control layer.\vspace{0.04in}

\noindent For all $t \geq \tau \geq 0$, the total duration $\bar{\Xi}(h_m,t)$ of DoS can be upper-bounded as follows over each time interval $[h_m,t)$:
\begin{equation*}
	|\bar{\Xi}(h_m,t)|  \leq |\Xi(h_m,t)| + n(h_m,t) t_{dos}  \leq \pi_\star + \frac{t - h_m}{\tau_\star}
\end{equation*}
where the constants $\pi_\star$ and $ \tau_\star$ are obtained using Features~\ref{assumption:DoSFrequency} and~\ref{assumption:DoSDuration}:
\begin{equation*}
	\pi_\star = \pi_d + \pi_f t_{dos} ~~~~~ \text{and} ~~~~~ \tau_\star = \frac{\tau_f \tau_d}{\tau_f + \tau_d t_{dos}}.
\end{equation*} 
Thus, the following inequality holds:
\begin{equation*}
	\begin{array}{rl}
		-\rho_v \bar{\Theta}(h_m,t) + \rho_{dos} \bar{\Xi}(h_m,t) & \leq -\rho_v \Big( t - \pi_\star - \frac{t-h_m}{\tau_\star}\Big) + \rho_{dos} \Big( \pi_\star + \frac{t-h_m}{\tau_\star} \Big)\\
				&\leq - \underbrace{\big(\rho_v - \frac{\rho_v + \rho_{dos}}{\tau_\star} \big)t}_{\text{Time-dependent}} + \underbrace{(\rho_v + \rho_{dos}) (\pi_\star - \frac{h_m}{\tau_\star})}_{\text{The ``$m^{th}$ DoS"-dependent}}.
	\end{array}
\end{equation*}
For the first component in the right hand side of~\eqref{eqM:DoS:Thm:FinalV} which is associated to $[0,t)$, we find:
\begin{equation*}
	\exp^{-\rho_v \bar{\Theta}(0,t) + \rho_{dos} \bar{\Xi}(0,t) } \bar{V}(0) = b_{dos} \exp^{-\big(\rho_v - \frac{\rho_v + \rho_{dos}}{\tau_\star} \big)t} \bar{V}(0)
\end{equation*}
where
\begin{equation*}
	b_{dos} = \exp^{(\rho_v + \rho_{dos}) \pi_\star}.
\end{equation*}
Further, other than $b_2 \exp^{-\sigma t}$, the rest of the components in the right hand side of~\eqref{eqM:DoS:Thm:FinalV} can be rewritten as follows: 
\begin{equation*}
\begin{array}{rl}
	b_2'(h_m)  \exp^{-\rho_v \bar{\Theta}(h_m,t) + \rho_{dos} \bar{\Xi}(h_m,t)} 
	      & = b_2 \exp^{(\rho_v - \sigma)h_m} \exp^{(\rho_v + \rho_{dos}) (\pi_\star - \frac{h_m}{\tau_\star})}  \exp^{-\big(\rho_v - \frac{\rho_v + \rho_{dos}}{\tau_\star} \big)t}\\
	      & =  b_2 b_{dos} \exp^{(\rho_v - \sigma - \frac{\rho_v + \rho_{dos}}{\tau_\star})h_m} \exp^{-\big(\rho_v - \frac{\rho_v + \rho_{dos}}{\tau_\star} \big)t}.
\end{array}
\end{equation*}
Therefore, we reach to the following compact representation as an upper-bound of $\bar{V}$ in \eqref{eqM:DoS:Thm:FinalV}:
\begin{equation*}
	\begin{array}{rl}
		\bar{V}(t) \leq b_{dos} \exp^{-\big(\rho_v - \frac{\rho_v + \rho_{dos}}{\tau_\star} \big)t} \bar{V}(0) + b_2 \exp^{-\sigma t}
							+ \sum_{m = 1}^{n} b_2 b_{dos}\exp^{-( \sigma  - \rho_v + \frac{\rho_v + \rho_{dos}}{\tau_\star})h_m} \exp^{-\big(\rho_v - \frac{\rho_v + \rho_{dos}}{\tau_\star} \big)t}.
	\end{array}
\end{equation*}
We use the definition of $\bar{V}(t) = x^T(t) \bar{P} x(t)$ and Rayleigh–Ritz inequality, and find:
\begin{equation*}
\begin{array}{rl}
     \|x(t)\|^2 \leq \frac{1}{\lambda_{min}(\bar{P})} \Big( & \lambda_{max}(\bar{P}) b_{dos} \exp^{-\big(\rho_v - \frac{\rho_v + \rho_{dos}}{\tau_\star} \big)t} \|x(0)\|^2 \\
     & + \sum_{m = 1}^{n} b_2 b_{dos}\exp^{-( \sigma  - \rho_v + \frac{\rho_v + \rho_{dos}}{\tau_\star})h_m} \exp^{-\big(\rho_v - \frac{\rho_v + \rho_{dos}}{\tau_\star}\big)t}
     + b_2 \exp^{-\sigma t} \Big)
\end{array}
\end{equation*} 
where both $t$ and $h_m$ may increase in time. When the condition~\eqref{eq:UM:DoS:Thm:Condition} is satisfied, this latter inequality is sufficient to conclude exponential convergence of all state trajectories to the origin in the presence of DoS.\vspace{0.05in}

\noindent \underline{\textit{(Part~3)}} The broadcast of information would be periodic after the detection of a DoS attack, e.g., via a TCP-like protocol. Therefore, we only focus on the event-triggered case of ETS~\eqref{eq:ETS}. To start the proof of Zeno-freeness, we note that the following inequality is guaranteed by ETS~\eqref{eq:ETS}:
\begin{equation*}
	\|e_{vi}\|^2 = \|K_i e_i\|^2 \leq \kappa_{1i} \exp^{-\sigma t} + \kappa_{2i} \|x_i\|^2.
\end{equation*}
Also, based on \eqref{eq:UM:UpperBoundonNormX_iSquare}, it is straightforward to find:
\begin{equation}\label{eq:UM:Zeno:UpperBoundOnNormE_i}
	\|e_i\| \leq \sqrt{b_{3i}} \exp^{-\frac{1}{2}\sigma t}
\end{equation}
where $b_{3i} = \frac{\kappa_{1i}+ b_1 \kappa_{2i}}{\lambda_{min}(K_i^T K_i)}$. By the definition of $e_i = \hat{x}_i - x_i$, we further know that the $i^{th}$ agent's triggering error evolves according to the following dynamics:
\begin{IEEEeqnarray*}{rll}
		\dot{e}_i = -A_i x_i - B_{u_i} K_i (\sum_{j \in \mathcal{N}_i^c} a_{ij}^c (x_i - x_j) + s_i^c x_i)  - B_{u_i} K_i (\sum_{j \in \mathcal{N}_i^c} a_{ij}^c (e_i - e_j) + s_i^c e_i) - B_{f_i} f_i(z_i).
\end{IEEEeqnarray*}
We rewrite this equation as follows:
\begin{IEEEeqnarray*}{rll}
	\dot{e}_i = -A_{ei} x_i - (\mathcal{L}_{ii}^c + s_i^c) B_{u_i} K_i e_i + B_{u_i} K_i \sum_{j \in \mathcal{N}_i^c} a_{ij}^c (e_j + x_j ) - B_{f_i} f_i(y_i)
\end{IEEEeqnarray*}
in which $A_{ei} = A_i + (\mathcal{L}_{ii}^c + s_i^c) B_{u_i} K_i$. Taking the 2-norm of both sides, we find:
\begin{IEEEeqnarray*}{rll}
		\|\dot{e}_i\| \leq \|A_{ei}\| \|x_i\| + \|B_{u_i} K_i\| \sum_{j \in \mathcal{N}_i^c} a_{ij}^c (\|e_j\| +  \|x_j\| ) + (\mathcal{L}_{ii}^c + s_i^c) \|B_{u_i} K_i\| \|e_i\|  +  \| B_{f_i} \| \|g_i(z_i)\|.
\end{IEEEeqnarray*}
Accordingly, using the fact $\frac{d}{dt}(\|e_i\|) = \frac{d}{dt}(\sqrt {e_i^T e_i}) = \frac{e_i^T \dot{e}_i}{\|e_i\|} \leq \frac{\|e_i\| \|\dot{e}_i\|}{\|e_i\|} \leq \|\dot{e}_i\|$ together with an upper bound on $\|x_i\|$ in \eqref{eq:UM:UpperBoundonNormX_iSquare} and on $\|e_i\|$ in \eqref{eq:UM:Zeno:UpperBoundOnNormE_i}, we find:
\begin{equation*}
	\frac{d}{dt}(e_i(t)) \leq b_{4i} \exp^{-\frac{1}{2}\sigma t}
\end{equation*}
where $b_{4i} = \big(\|A_{ei}\| + \mathcal{L}_{ii}^c \|B_{u_i} K_i\| + \|B_{f_i}\|\|\mathcal{A}_a\| \sqrt{\gamma_f \gamma_{cz}}\big) \sqrt{b_1} + \big( \sum_{j \in \mathcal{N}_i^c} a_{ij}^c \sqrt{b_{2j}} + (\mathcal{L}_{ii}^c + s_i^c) \sqrt{b_{3i}} \big) \|B_{u_i}K_i\|$ is a positive scalar. We integrate both sides of the above inequality over $t \in [t_k^i, t_{k+1}^i)$, and use the fact $e_i(t_{k+1}^i) = \textbf{0}$ (guaranteed by ETS~\eqref{eq:ETS}) together with the comparison lemma in order to reach to a new upper bound on the norm of triggering error:
\begin{equation*}
	\|e_i (t)\| \leq \frac{2 b_{4i}}{\sigma} (\exp^{\frac{1}{2} \sigma t_k^i} - \exp^{-\frac{1}{2} \sigma t}).
\end{equation*}
We use ETS~\eqref{eq:ETS} to further lower bound the above inequality at the (next) triggering time $t_{k+1}^i$:
\begin{equation*}
	\sqrt{\kappa_{1i}} \exp^{-\frac{1}{2} \sigma t_{k+1}^i} \leq \|e_i(t_{k+1}^i)\| \leq \frac{2 b_{4i}}{\sigma} (\exp^{\frac{1}{2} \sigma t_k^i} - \exp^{-\frac{1}{2} \sigma t_{k+1}^i}).
\end{equation*}
Therefore, based on the (lower and upper) bounds on the left and right hand sides, we reach to:
\begin{equation}\label{eq:UM:Zeno:FinalLowerBound}
	t_{k+1}^i - t_k^i \geq \frac{2}{\sigma} \ln \big(1 + \frac{\sigma \sqrt{\kappa_{1i}}}{2b_{4i}} \big)> 0.
\end{equation}
This strictly positive lower bound on $t_{k+1}^i - t_k^i$ guarantees Zeno-freeness for the proposed ETS-based distributed stabilization protocol in the absence of DoS.\vspace{0.05in}
\end{proof}


\begin{thebibliography}{}
	\bibitem{Khalil-Prentice-2003} Khalil H., \emph{Nonlinear Systems}, Prentice-Hall, 2003.	
\end{thebibliography}
\end{document}